\documentstyle[12pt,aas2pp4,epsfig]{article}

\slugcomment{}

\lefthead{}
\righthead{}

\begin{document}

\title{Power Spectrum of the density of cold atomic gas \\ 
       in the Galaxy towards Cas A and Cygnus A}

\author{A. A. Deshpande, K. S. Dwarakanath}
\affil{Raman Research Institute, Bangalore 560 080, India}

\and

\author{W. M. Goss}
\affil{National Radio Astronomy Observatory, PO Box O, Socorro, NM 87801}

\begin{abstract}

We have obtained the power spectral description of the density and opacity fluctuations 
of the cold HI gas in
the Galaxy towards Cas A, and Cygnus A. 
We have employed a method of deconvolution, based on CLEAN, 
to estimate the true power spectrum of optical depth of cold HI gas from the observed distribution, 
taking into account the finite extent
of the background source and the incomplete sampling of optical depth over the extent of the source.
We investigate the nature of the underlying spectrum of density fluctuations
in the cold HI gas which would be consistent with that of the observed HI optical depth fluctuations. 
These power spectra
for the Perseus arm towards Cas A, and for the Outer arm towards
Cygnus A have a slope of 2.75$\pm$0.25 (3$\sigma$ error). The slope in the case of the 
Local arm towards Cygnus A is 2.5, and is significantly shallower in comparison. 
The linear scales probed here range from 0.01 to 3 pc. We discuss the implications of our 
results, the non-Kolmogorov
nature of the spectrum, and the observed HI opacity variations on small transverse scales.

\end{abstract}


\section{INTRODUCTION}

The cold atomic gas in the Galaxy has been observed for over forty years in the 
21 cm HI line. These early single dish and interferometric observations established 
the presence of cold atomic gas in the Galaxy possibly in the form of discrete  
$'$clouds$'$ in pressure equilibrium with the warm neutral medium (Clark, Radhakrishnan,
\& Wilson 1962, Clark 1965, Radhakrishnan et al. 1972, Field, Goldsmith, \& Habing 1969,
Field 1973). These HI clouds with an estimated mean spin temperature of $\sim$ 80 K, and a 
mean density of $\sim$ 30 cm$^{-3}$  were thought to be a few parsec in size. This simple
picture of the cold gas has changed significantly during the last two decades due to high resolution 
21 cm line absorption measurements of the Galaxy made using aperture synthesis telescopes (Greisen
1973,  Lockhart, \& Goss 1978, Bregman et al. 1983, Schwarz et al. 1986, 1995).
Fine structures in the atomic hydrogen gas 
have been observed over a wide range in length scales down to about 0.1 pc
(Crovisier et al. 1985, Kalberla et al. 1985, Bieging et al. 1991, Roberts 
et al. 1993, Schwarz et al. 1995). Some lines of sight which have also been explored
through VLBI and multi-epoch pulsar observations have revealed 
significant opacity variations 
in the atomic gas on transverse scales down to $\sim$ 10 AU (Dieter et al. 1976, 
Diamond et al. 1989,
Deshpande et al. 1992, Frail et al. 1994, Davis et al. 1996, Heiles 1997, Faison et al. 1998). 
The observed opacity variations had raised concerns regarding the origin, lifetime, and
pressure equilibrium with the rest of the interstellar medium of small scale structure 
presumed to cause them. Related issues in this context are
coherent length scales of the underlying structures, the role of turbulence in producing these, and the
power contained in such structures as opposed to that in large scale structures.

Some of these issues are best addressed through a power spectrum analysis of the HI density 
fluctuations at the relevant range of scales. For the warm HI gas observed in emission, this is rather
straightforward since the observed 21 cm line intensity is directly proportional to the
column density of the HI gas. Thus, the observed line visibilities as a function of spatial
frequency define
the power spectrum of the HI density fluctuations. For scales of 50 to 200 pc this has
been carried out by Green (1993), and a useful analytical formulation related to such 
measurements is discussed by Lazarian (1995). The measurement of HI absorption distribution, 
on the other hand, can not be directly interpreted in terms of 
the power spectrum of the HI density fluctuations, as 
the HI absorption depends on the integral along the sight line of the ratio of 
the number density to the spin temperature of the HI gas. Thus, the translation of the observed
optical depth fluctuations to those in HI density necessarily involves an assumption about the
dependence of the density on temperature (e.g., pressure equilibrium would imply constancy of
the product of density and temperature ). Direct interpretations of the HI absorption measurements
are made difficult by the effects due to the finite extent and the shape of the background source.

More recently, the opacity variations in the cold atomic gas have been imaged using bright
supernova remnants as background continuum sources (Beiging, Goss, \& Wilcots 1991, Roberts
et al 1993, Schwarz et al 1995). Some of these observations provide a unique opportunity
to study the opacity distributions over a large range of spatial scales. 
In this paper, we attempt to derive an underlying power spectrum from
the HI opacity distribution observed in the direction of two bright radio sources, Cas A, 
and Cygnus A. The power spectrum of the opacity distribution over about two orders of magnitude
in spatial scales (as from the Cas A data) is found to follow a power law.  
The optical depth images used in the current analysis are discussed in Section 2. The power spectrum
analysis and application of CLEAN for high-dynamic-range spectral estimation are discussed in 
Section 3. This CLEAN procedure, to recover the true spectrum from the observed spectrum, 
takes into account the extent and shape of the background source. In section 4, the estimated 
power spectra 
of the cold HI gas in the Galaxy towards Cas A are presented. In this section, we also
present the structure function of the observed optical depth distribution in the Galaxy 
towards the direction of both the sources.
The possible relation between the opacity spectrum and the underlying spectrum of HI density 
fluctuations is examined in Section 4.1. The implications of the derived power spectra of the
cold gas for the interstellar medium are discussed in Section 5.

\section{OPTICAL DEPTH IMAGES}

The distribution of cold HI gas in the Galaxy towards the supernova remnant 
Cassiopeia A ($l_{II}\sim 111^{o}, b_{II}\sim -2^{o}$) was observed by 
Bieging, Goss, \& Wilcots (1991; hereafter BGW) using the Very 
Large Array (VLA) of the National Radio Astronomy Observatory. They present optical depth 
images at an angular resolution
of 7$''$ and a spectral resolution of 0.6 km s$^{-1}$. 
Cas A is at a distance of $\sim$ 3 kpc from the Sun and 
is believed to be in, or, beyond the Perseus spiral arm. 
The optical-depth image cube presented by BGW covers a
velocity range from  $\sim$ --30 km s$^{-1}$ to --56 km s$^{-1}$ (V$_{lsr}$). The HI
absorption in this velocity range is caused by the interstellar gas in the Perseus spiral arm.
An optical depth image in one of the velocity channels (at V$_{lsr}$ = -38.8 km s$^{-1}$)
 from this cube is reproduced in Fig. 1. 
The morphology of the absorbing gas at an adopted distance of $\sim$ 2 kpc to the Perseus arm
(spatial resolution $\sim$ 0.07 pc), 
is dominated by filaments, loops, and arcs. 
\begin{figure*}
\epsfig{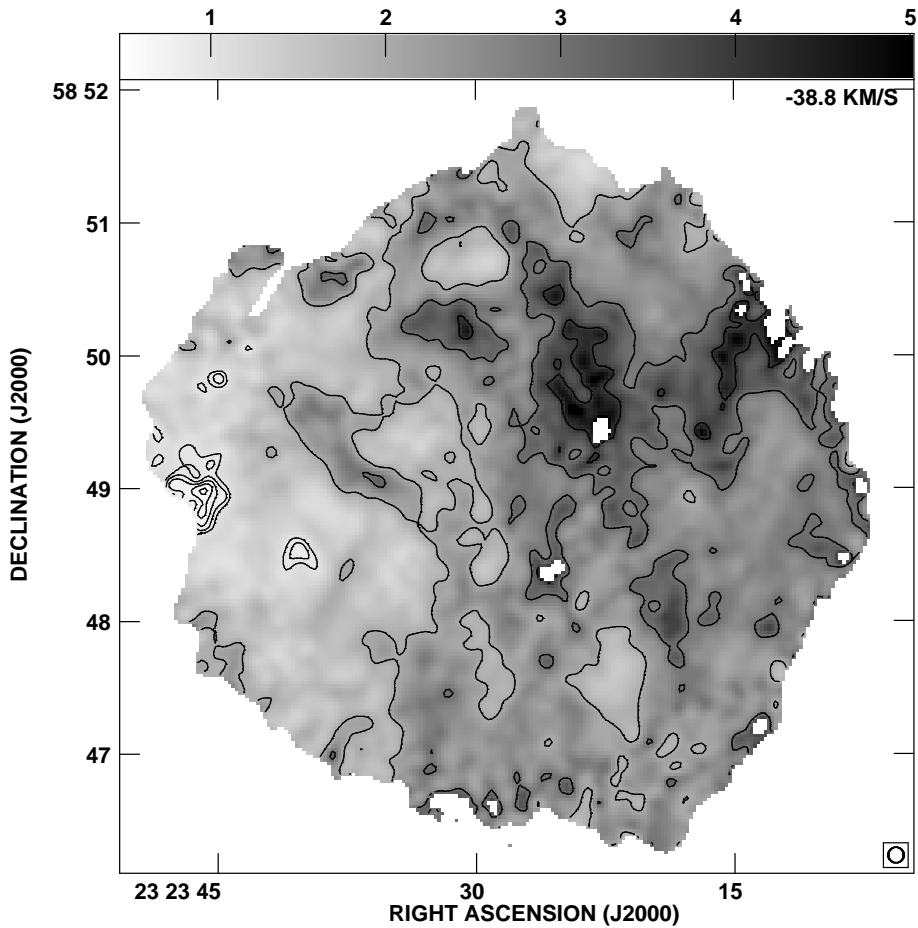}
\caption{ The HI optical depth image at a V$_{lsr} = -$38.8 km s$^{-1}$
towards Cas A from BGW. This VLA image has an angular resolution of 7$''$ with a sampling interval of
1$''$.6. The optical depth grey scale 
ranges from 0.5(white) to 5(black). The contours are in units of 0.1 from 0.1 to 1, 
and in units of 1 from 1 to 5 in optical depth. The 
synthesized beam is shown in the bottom right hand corner. The velocity resolution of
this image is 0.6 km s$^{-1}$.}
\end{figure*}
Recently, we used the VLA to observe the  
distribution of cold gas in the Galaxy
toward the direction of the powerful extragalactic radio source Cygnus A ($l_{II}
\sim 75^{o}, b_{II}\sim 5^{o}$). The observations were carried out with the VLA in
the B configuration (maximum baseline $\sim$ 11 km) on December 26, and 27, 1999. A bandwidth of
1.56 MHz was used with 256 channels, and Hanning smoothing, resulting in a 
spectral resolution of 1.29 km s$^{-1}$. The synthesized beam was 3$''$.8 $\times$ 
3$''$.6 at a position angle of --2$^{o}$.8. Bandpass calibration was achieved by 
frequency switching on Cygnus A. 
The channels from the flat portion of the spectrum and free from line absorption 
were used to estimate
continuum visibilities. Spectral-line visibilities in each of the relevant spectral channels 
were obtained by removing the estimated continuum contribution to the visibilities.
The images constituting a spectral image cube, as well as the continuum image were
constructed and CLEANed.
The rms noise in each of the spectral channels and in the continuum image was 0.1 Jy/beam,
limited by dynamic range.
The optical depths ($\tau$) were estimated in the standard way at all positions where 
the line intensity and the continuum intensity were greater than five times the rms
value in the respective images. Some examples of optical depth images 
produced this way are shown in Figs. 2, and 3. 
The double-lobed morphology is simply a reflection of the 
radio continuum morphology of Cygnus A.
The optical
depth images in Figs. 2, and 3 provide examples of the distribution of cold atomic gas in the Local, 
and the Outer spiral arms, respectively. The spatial resolutions of these images
correspond to $\sim$ 0.01 pc, and 0.3 pc at the Local, and the Outer arms respectively. The assumed
distances to the Local, and the Outer arms are 0.5, and 15 kpc, respectively. There is very little
HI absorption (peak optical depth $<$ 0.05) in the Perseus arm towards Cygnus A (Fig. 4). Consequently, 
except towards the two brightest positions in the two lobes of Cygnus A, the optical depth of the
Perseus arm remains
unmeasured towards most of the area covered by the two lobes. These images are not displayed.
For the same reason, we also do not use the Perseus arm optical depth images towards Cygnus A 
in the structure function analysis discussed in Section 4.    
\begin{figure*}
\epsfig{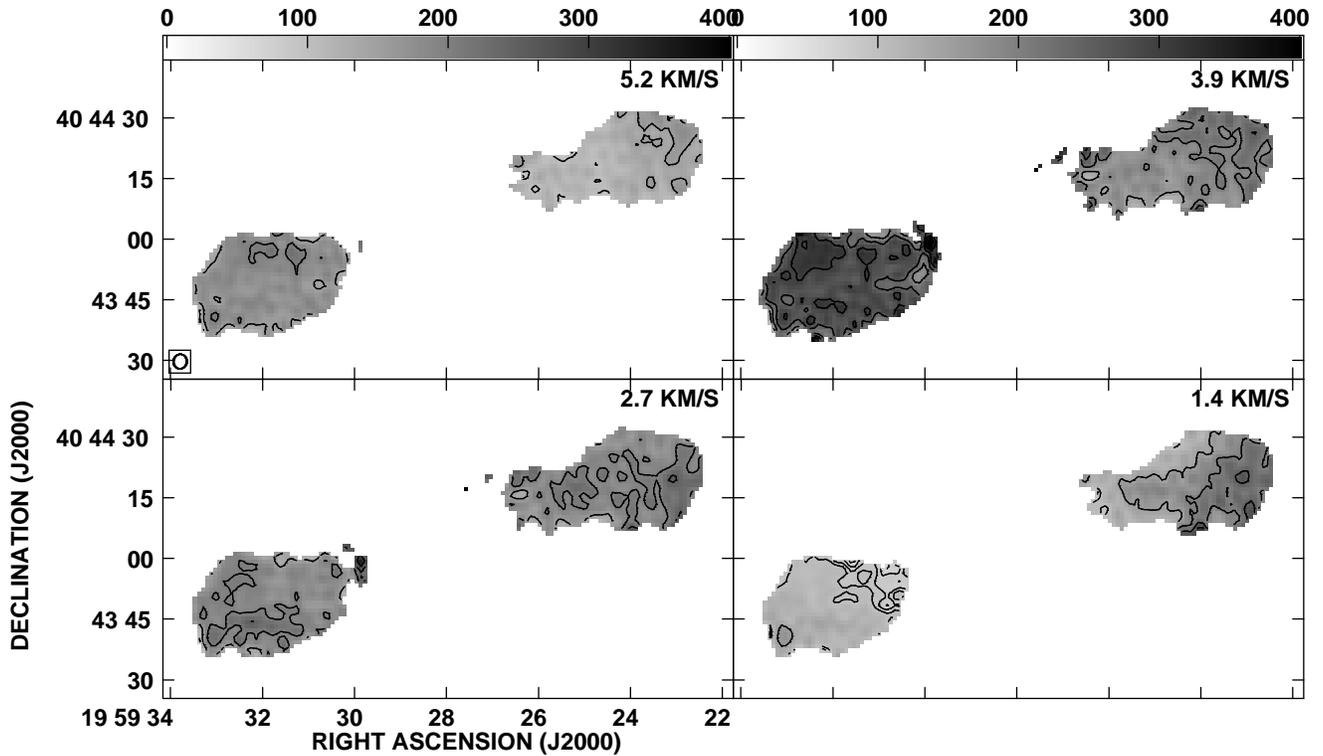}
\caption{The HI optical depth image of the Local arm towards Cygnus A produced 
from the VLA observations described in this paper. The top
right hand corner indicates the V$_{lsr}$. The bottom left hand corner of the first panel 
shows the synthesized
beam. The optical depth grey scale ranges from 0 (white) to 0.4 (black). The contours are in units of
0.01 from 0.01 to 0.1, and in units of 0.05 from 0.1 to 0.5
in optical depth. The angular resolution of $\sim$ 3.$^{''}$7 corresponds to a linear 
resolution $\sim$ 0.01 pc for an assumed distance of 0.5 kpc to the Local arm.  
The sampling interval in these images is 1$''$, and the velocity resolution is 1.29 km s$^{-1}$.}
\end{figure*}
\begin{figure*}
\epsfig{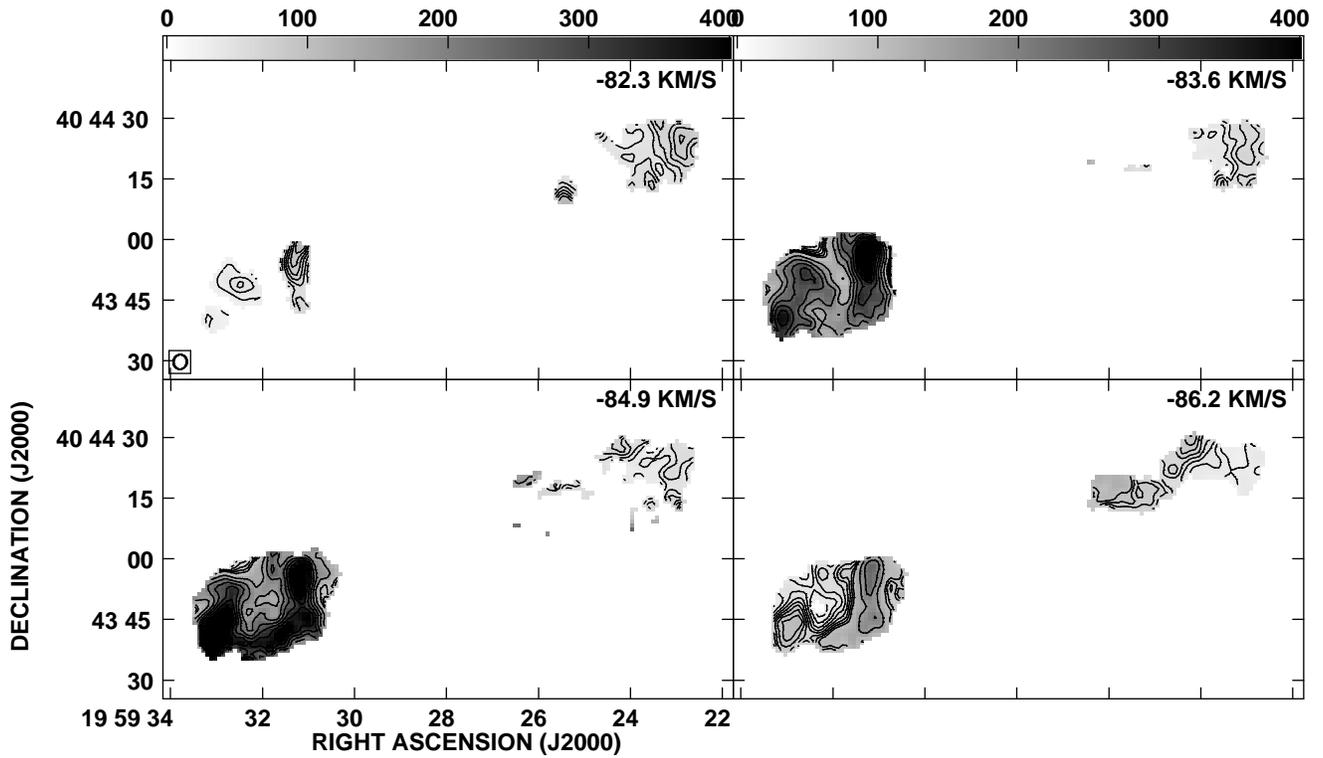}
\caption{The HI optical depth image of the Outer arm towards Cygnus A obtained from the
VLA. The contours are
as in Fig. 2. 
The angular resolution of $\sim$ 3.$^{''}$7 corresponds to a linear size $\sim$ 0.27 pc for an 
assumed distance
of 15 kpc to the Outer arm.}
\end{figure*}
\begin{figure*}
\epsfig{file=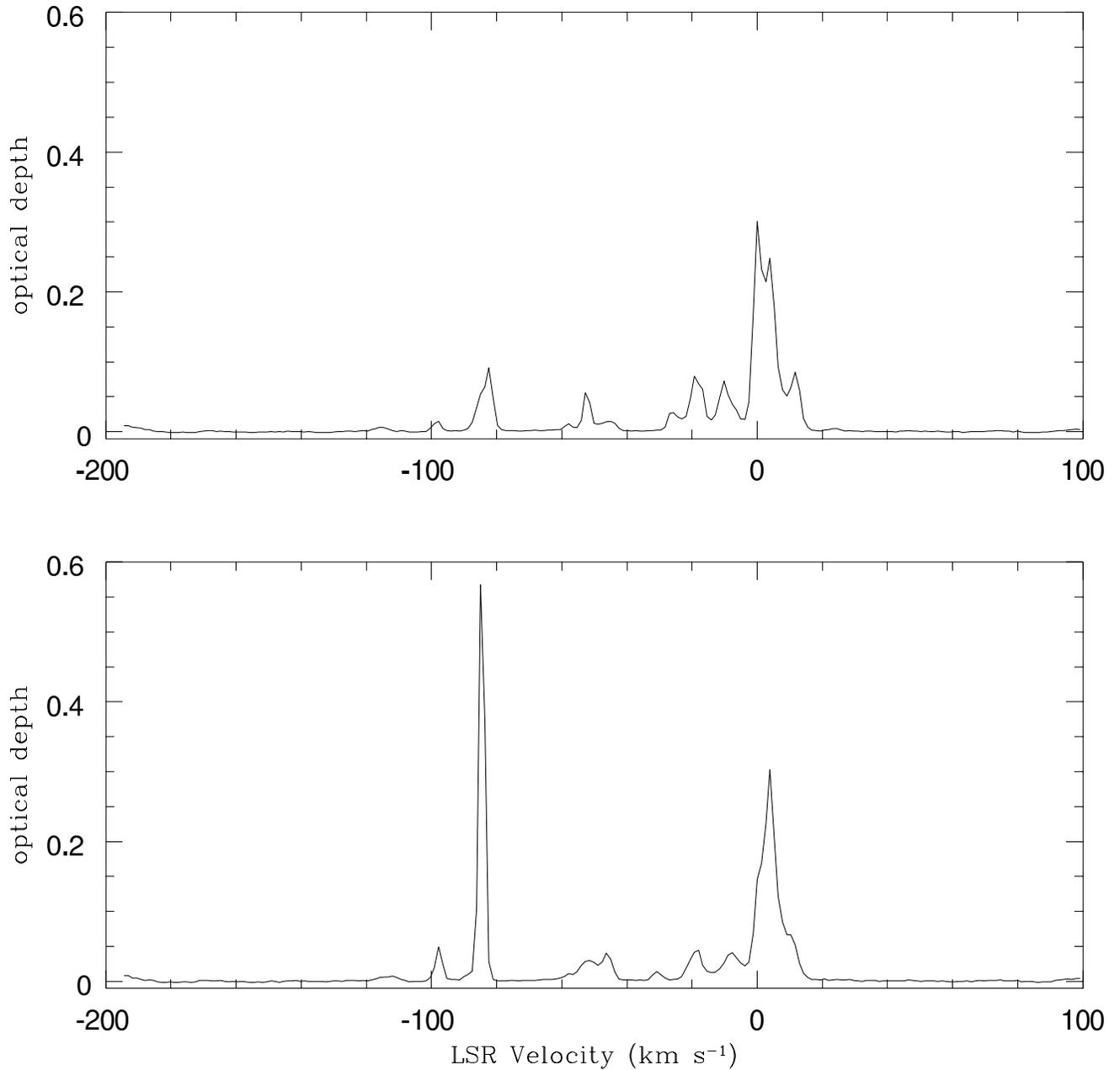, height=\textwidth}
\caption{ The upper and lower frames show the HI optical depth spectra towards the northwest 
and the southeast peaks of Cygnus A respectively. Note the large optical depth values corresponding
to the Local (V$_{lsr} \sim$ 0 km s$^{-1}$), and the Outer (V$_{lsr} \sim$ --85 km s$^{-1}$) arms.
In contrast, the Perseus arm shows a weak HI absorption ($\tau <$ 0.05) at V$_{lsr} \sim$ --50 km s$^{-1}$.}
\end{figure*}
\section{POWER SPECTRUM ANALYSIS}

For the first time, the optical depth images in the Cas A direction provide possibilities
of studying the distribution of cold HI over a usefully wide range of  spatial
scales simultaneously. Given the extent and the resolution of these images, a detailed estimate of the
power spectrum of the opacity distribution over nearly two decades in spatial scales is 
possible.  In this section, we attempt such an estimate and present details of our analysis.

In an ideal situation where a fully sampled opacity distribution is available over an
extent of $\theta_{max}$  and with a sampling interval $\Delta \theta$, it is
possible to simply Fourier transform the distribution and estimate the power spectrum
over a frequency range 1/$\theta_{max}$ to 1/(2$\Delta\theta$). 
In reality, the opacity distribution is not fully sampled (see Fig. 1), 
making it difficult to estimate the true power-spectrum  directly. As is clear from 
the distribution in Fig. 1, the incomplete sampling is
primarily due to the finite extent of the background source, Cas A. In principle,
 a smaller section of the image where sampling is complete can be chosen; however,
reduced resolution in the spectrum then results.  Unfortunately, some incompleteness 
exists in most of the optical depth images within the source extent where the 
continuum background or 
the line intensity is too weak to allow a reliable estimation of the optical depth. 
If the optical depth at the unsampled locations in the image is assigned `zero' value,
the Fourier transform of the image gives a spectrum that is a convolution of the
true spectrum with the Fourier transform of the measurement 
mask or the {\it sampling spectrum}.  The measurement mask
is defined by ones and zeros at the locations that are sampled, and unsampled, respectively.
This convolution can significantly modify the spectrum, 
thus making the nature of the true spectrum less apparent. Some form of deconvolution 
operation is therefore required
to {\it recover} the true spectrum, particularly when high dynamic range estimation is 
a primary consideration (as would be while studying steep power-law spectra). 

\subsection{Application of CLEAN for high-dynamic-range spectral estimation}

The situation in obtaining the `true' spectrum from the modified or {\it dirty} 
spectrum is analogous 
to that of obtaining a clean image from a dirty image routinely encountered 
in synthesis imaging (Ryle 1960). 
There, the `dirty image' obtained
from a telescope is the convolution of the `clean image' with the `point spread function'
(PSF) of the telescope and the dirty image needs to be deconvolved with the PSF in 
order to obtain the true (or clean) image. The deconvolution attains even more importance when 
the telescope is an unfilled aperture. 
In such a situation a technique called CLEAN (H\"ogbom 1974) is commonly used to obtain 
high-dynamic-range images. 
In the present situation, the dirty spectrum is to be deconvolved using the 
spectrum of the mask in order to obtain the true spectrum.
The application of CLEAN
 in the power spectral analysis of non-uniformly sampled pulsar timing data
was developed by Deshpande et al. (1996). In that case a 1-dimensional CLEAN was developed.
In the present context a 2-dimensional variant of this application is used. Details of the application
of CLEAN in the context of power spectral analysis can be obtained from Deshpande et al. (1996).
Only a brief summary of the procedure used in the present context will be given here.

The 2-dimensional $\tau$ image (for e.g., Fig. 1) which is 223 $\times$ 223 pixels is padded
 with zeros to obtain a 512 $\times$ 512 image. A mask of the same dimensions is also prepared. 
This padding helps
in improving the sampling by a factor of $\sim$2 in the spectral domain which improves the 
performance of CLEAN qualitatively, although at the expensive additional computation.
Both the image and the mask in the image domain are Fourier transformed to obtain the image 
spectrum and the mask spectrum in the spatial-frequency domain, respectively. 
The spectra so obtained
are hermitian symmetric (as are visibilities and unlike the images in the case of synthesis imaging) 
and the CLEAN algorithm was modified to account for this modification: (i) During each iteration
of CLEAN, while searching for the maximum in the spectrum, 
 only the amplitudes were considered. But, while removing CLEAN components, the 
complex nature (amplitude, and phase)
is  taken into account, and, (ii) the search for the maximum was made only over one half of the image 
spectrum. Subsequently, subtraction of a scaled version of the mask spectrum was 
performed over both halves
of the image spectrum after accounting for the hermitian symmetric contribution. This procedure ensures 
the hermitian symmetry in the CLEAN spectrum. The CLEAN complex components are restored with 
a half-cycle cosine of  half power width close to the original spectral resolution which is  
added to the residuals. The square of the amplitude of the resultant CLEAN complex spectrum 
is then used to estimate the power spectrum.
Finally, an azimuthally-averaged 1-d power spectrum is obtained from such a 2-d CLEAN power spectrum 
to improve the spectral definition; of course any information on anisotropy is lost.
This procedure was applied to a series of optical-depth images corresponding to individual velocity
channels as well as those integrated over wider velocity ranges.
Given the continuous nature of the spectra, one may legitimately
question the applicability of the CLEAN deconvolution procedure (Schwarz 1978), since 
the CLEAN assumes a generally
`empty' domain with only well separated narrow spectral components in the `true' spectrum. 
Interestingly, 
the often encountered steep power-law spectra (power $\propto q^{-\alpha}$, where q is the spatial
 frequency) are `red' in nature, 
i.e., $\alpha$ is positive. When viewed on a linear-linear
scale, such spectra would be highly peaked around `zero' frequency and relatively `empty' elsewhere,  
providing a favorable situation for the CLEAN.

The azimuthally-averaged mask spectrum corresponding to the image
in Fig. 1 is a featureless power-law (power $\propto$ q$^{-\alpha}$, where q is the spatial 
frequency) with a slope (or power-law index)
of 2.9 (=$\alpha$). This spectrum can be understood by noting that
the optical-depth image in Fig. 1 defines an approximately circular area (the 
morphology of the background source Cas A) as the sampled distribution. 
The Fourier transform of a uniformly weighted perfect circular patch is an Airy pattern, i.e.,  
a Bessel function of I order divided by q. For large values of its argument (q), 
the square of the  Airy pattern decreases as $q^{-3}$. 
Hence, the observed power-law slope of the mask-spectrum is not surprising.
As an example of the improvement brought about by the CLEAN procedure, Fig. 5 shows the 
azimuthally-averaged versions of the dirty-spectrum (i.e. pre-CLEAN, shown as dashed line) 
and the clean-spectrum 
(i.e. CLEANed and restored, shown as solid line) for the images similar to that in Fig. 1.
In both cases the power spectra from eleven channels in the velocity range
--41.4 km s$^{-1}$ to --34.9 km s$^{-1}$ were averaged to improve the spectral definition, 
since the individual noisy spectra are similar.
There are two distinct differences between the two spectra.
First, the {\it true} spectrum has a knee at log(q/q$_{max}$) $\sim -$0.5 corresponding to the size of the 
synthesized beam $\sim$ 7$''$.
The power spectrum drops sharply beyond these frequencies. This `bowl' is filled in by the `leakage'
of power from lower frequencies in the dirty spectrum. Second, the slope of the true spectrum is 
2.75 while that of the dirty spectrum is 2.4 (both the slopes calculated between
--0.5 and --1.5 in log(q/q$_{max}$)).
The dirty spectrum is expected to be flatter than
the CLEAN spectrum since the latter has been convolved with the mask spectrum leading to a 
mixing (spread) of power from smaller to larger spatial-frequencies.

\begin{figure*}
\epsfig{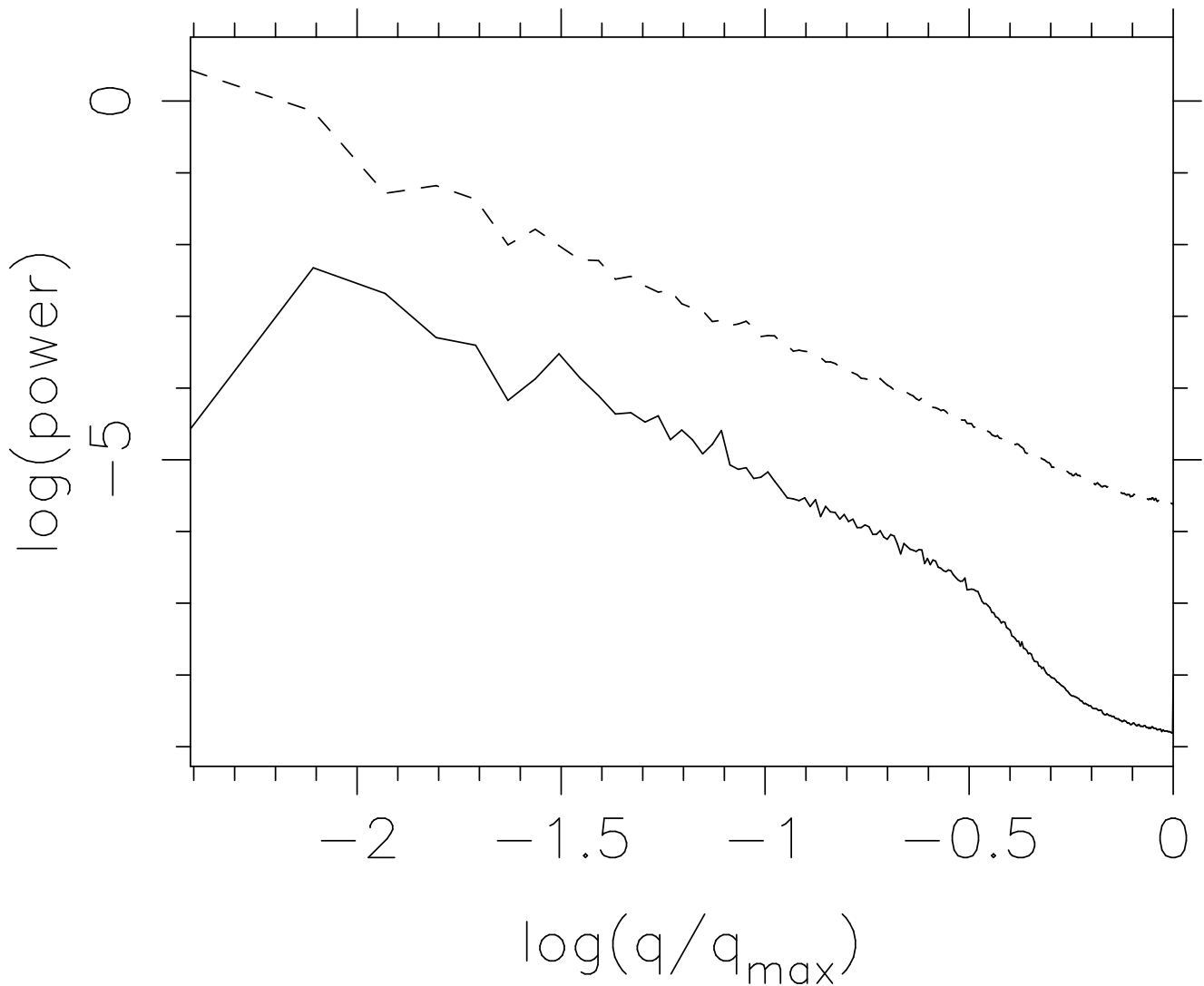}
\caption{Demonstration of the CLEAN  procedure in power spectra. The x-ordinate is 
logarithm of the spatial frequency in units of the limiting frequency (q$_{max}$). The limiting
frequency is half of the sampling frequency.  The sampling interval in the
Cas A optical depth images used here is 1.$''$6.  The top curve (dashed line) is a dirty
power spectrum obtained after averaging the power spectra of 11 channels 
 in the velocity range V$_{lsr}$ = --41.4 km s$^{-1}$ to --34.9 km s$^{-1}$. The bottom curve 
(solid line) is the true power spectrum obtained
after CLEANing. Note the appearance of the $'$knee$'$ in this spectrum at log(q/q$_{max}$)
$\sim$--0.5. This effect is due to the finite
size (7$''$) of the synthesized beam. The slopes of the dirty and true spectra are 2.4 and 2.75 
respectively, estimated over the x-range --0.5 to --1.5.}
\end{figure*}

It may be noted that although the processing described so far involves 2-dimensional images/spectra, 
the results are presented as azimuthally averaged versions. This averaging improves the definition 
of the spectra in a monotonically increasing fashion towards higher spatial frequencies consistent
with the increasing number of independent samples available for the averaging. Thus the uncertainties
are relatively high towards the lower spatial frequency end of the spectrum, which is therefore 
excluded from any quantitative estimation (e.g. slope). At the other end,  
the shape of the $'$bowl$'$ is a result of the steepening of the spectrum due to beam smoothing
and a flattening of the spectrum as the contribution due to the uncertainties in the
estimation of optical depth becomes significant. 
Consideration of these two effects
defines the range of spatial frequencies over which the spectral characteristics can be estimated 
reliably. The CLEAN spectrum viewed over this range is describable by a simple power law of the
nature q$^{-\alpha}$, where q is the spatial frequency, and $\alpha$ is the power-law index.

\section{POWER SPECTRA FROM CAS A AND CYGNUS A}

The spatial power spectra of optical-depth distribution towards Cas A are shown in Fig. 6.
These spectra represent average description over the velocity intervals --51.6 to --45.2 km s$^{-1}$ 
and --41.4 to --34.9 km s$^{-1}$ separately. These velocity ranges correspond to the Perseus arm and 
 arise from a distance of 2 kpc.
In Fig. 6 there are three spectra and are obtained
as follows : (1) In the first case (Case A), a power spectrum is obtained 
for each of the 11 
velocity channels in the chosen velocity range (for e.g., --51.6 to --45.2 km s$^{-1}$). 
The resultant 11 power spectra were found to be similar in slope and differing in the spectral
density consistent with the optical depth variations in the images used.
They were combined with uniform weighting giving an average description of opacity distribution
viewed over a narrow channel (width 0.6 km s$^{-1}$ in the present case). In Fig. 6 there are two
spectra obtained in this fashion corresponding to the two velocity ranges --51.6 to --45.2 km s$^{-1}$
(indicated by dots and dashes) and --41.4 to --34.9 km s$^{-1}$ (solid line). 
These two spectra are less noisy (owing to the averaging over 11 channels) and follow each other 
well in the log(q/q$_{max}$) range of --1.5 to --0.5. 
It is not surprising that the agreement of the power spectra in
these two velocity ranges is excellent given the same range of spatial scales covered.
In the second case (Case B), the optical depth images over 11 velocity channels of a chosen 
velocity range (--41.4 to --34.9 km s$^{-1}$) were first averaged to obtain one image for 
which a power spectrum was estimated. 
A given pixel in the average image was blanked when the average could not be performed over
all the 11 velocity channels (i.e., when one or more of the channel images had the corresponding 
pixel  blanked). This reduces, however, the total number of $'$valid$'$ pixels in the resultant
image. This effect was minimized for the velocity range --41.4 to --34.9 km s$^{-1}$.
This spectrum is shown in Fig. 6 as a dashed-line  and
is easily recognized as the noisiest of the three spectra shown. The slopes estimated over the
log(q/q$_{max}$) range of --0.5 and --1.5 for all the three spectra are equal  within the
noise, with a power law slope of 2.75$\pm$0.25 (3$\sigma$ error).  The average power in the 
Case B spectrum (dashed line) over this range in log(q/q$_{max}$) is about a factor of three
higher than the two Case A spectra. The implications of this enhanced
power will be discussed in Section 5. 

\begin{figure*}
\epsfig{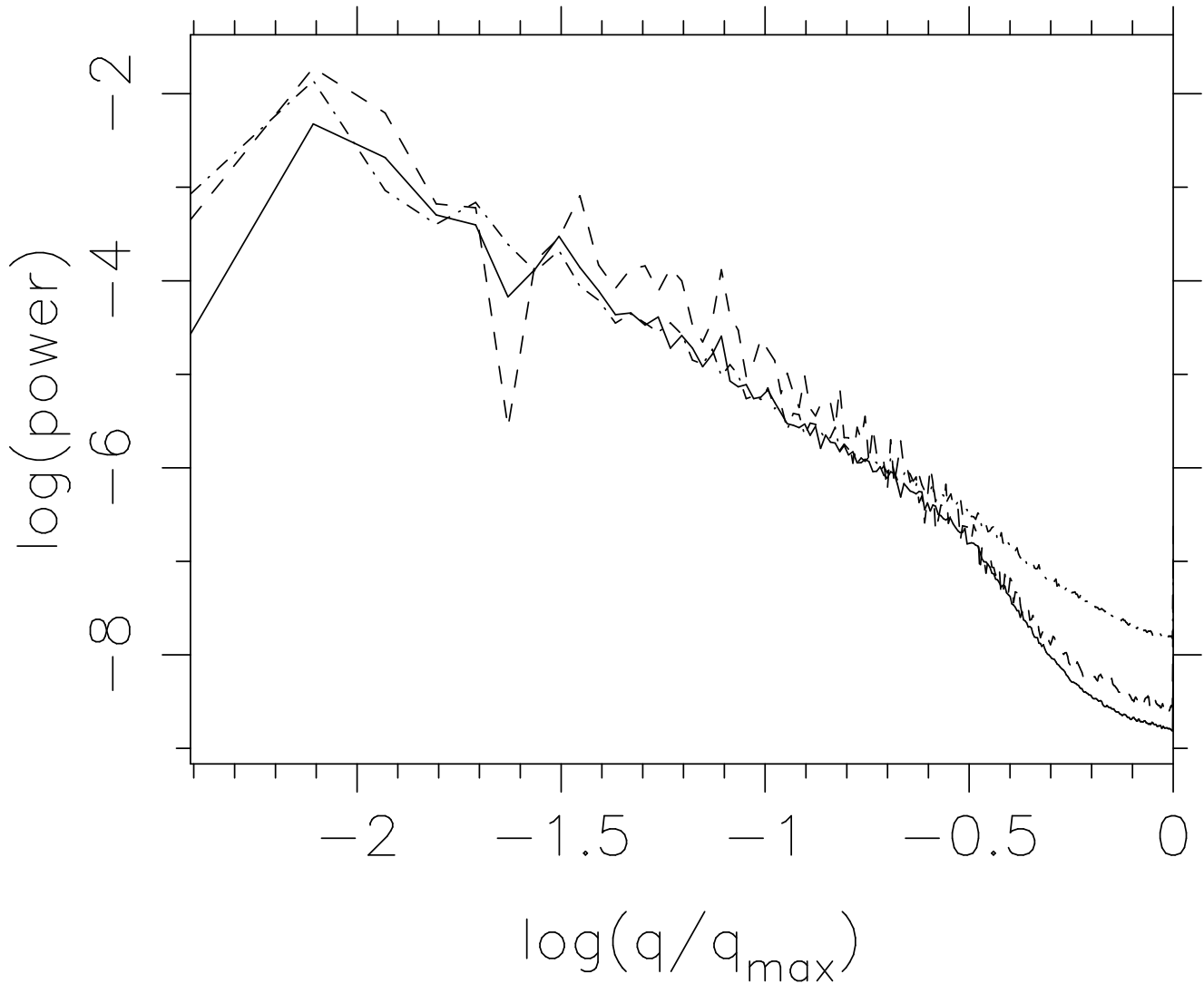}
\caption{The spatial power spectra of optical depth distribution towards Cas A. The
spectrum indicated by the solid line is obtained as in Case A (Section 4) and corresponds
to the velocity range --41.4 to --34.9 km s$^{-1}$. The spectrum indicated by the 
dot-dashed line also corresponds to Case A, but to the velocity range --51.6 to --45.2 km s$^{-1}$.
The third spectrum (dashed line) is obtained as in Case B (see Section 4) and corresponds to
the velocity range --41.4 to --34.9 km s$^{-1}$. 
All the three spectra have a slope of 2.75$\pm$0.25 (3$\sigma$ error) over
the log(q/q$_{max}$) range of --0.5 and --1.5. However, the Case B spectrum has three times higher power
than the Case A spectrum  over this range of log(q/q$_{max}$).}
\end{figure*}

A similar analysis was also attempted for the data in the direction of Cygnus A . However, given the
very limited sampling in this case, a structure function should provide a more useful
description of the opacity distribution than an azimuthally averaged spectrum even after
CLEAN.  Hence, we have obtained the structure function for the observed optical depth 
distribution
towards Cygnus A and compared this with the Cas A results. The structure function  
of the optical depth is defined as $D_{\tau}(r) = <[\tau (x) - \tau (x-r)]^{2}>$, where
$\tau (x)$ is the optical depth at point $x$, and the estimation is based on only 
those points in the image where the optical depth estimates are considered to be reliable.
It can be shown that for a distribution having a power law spectrum ($\propto q^{-\alpha}$),
the structure function will also be a power law ($\propto r^{2\beta}$, where $r \propto 1/q$), 
such that $\alpha$ = 2+2$\beta$ (Lee \& Jokipii 1975). 

In Fig. 7, the square root of the structure function (i.e., the rms fluctuation) of the optical depth 
distribution in the Perseus arm towards Cas A (indicated by $\times$s) is compared with 
those in the Local (indicated by circles) and 
Outer arms (indicated by squares) towards Cygnus A. The spatial scales indicated are 
appropriate to the assumed
 distances of the absorbing HI gas.  The effect of the beam smoothing is
apparent towards the smaller spatial scales (over about half an order of magnitude) in each
of the structure functions. Towards the larger spatial scales, 
the estimation suffers from an inherent limitation in having a limited number of independent realizations
of the large scale variations that are required to obtain a true average estimate. The gap 
seen in the structure function corresponding to the Outer arm of Cygnus A results from inadequate statistics
consistent with the absence of optical depth estimates over a large fraction of the north-west
lobe (see Fig. 3). Even with these limitations the reliably estimated sections of the  
structure functions indicate that the slopes of two of the three determinations (the Perseus arm,
and the Outer arm)  are consistent ($\beta \sim$0.4). In addition, this value of $\beta$ is
consistent  with the power law index ($\alpha$ = 2.75) for the gas in the 
Perseus arm (Fig. 6). The slope in the case of the 
Local arm is significantly shallower ($\beta \sim$0.25). In addition, the
fluctuations in the opacity (and its mean value) are an order of magnitude higher for the 
gas associated with the Perseus arm of Cas A as compared to the 
Outer arm HI in the direction of Cygnus A. The vertical heights from the Galactic plane of 
the HI gas observed in these 
two cases are different, viz., $\sim$ 70 pc for the gas in the Perseus arm towards Cas A,
and $\sim$ 1300 pc for the gas in the Outer arm towards Cygnus A.  However, it is unclear how much 
of the differences seen in the structure functions corresponding to these arms are 
attributable to the differences in the heights from the Galactic plane of the absorbing
HI gas in these two arms. 

\begin{figure*}
\epsfig{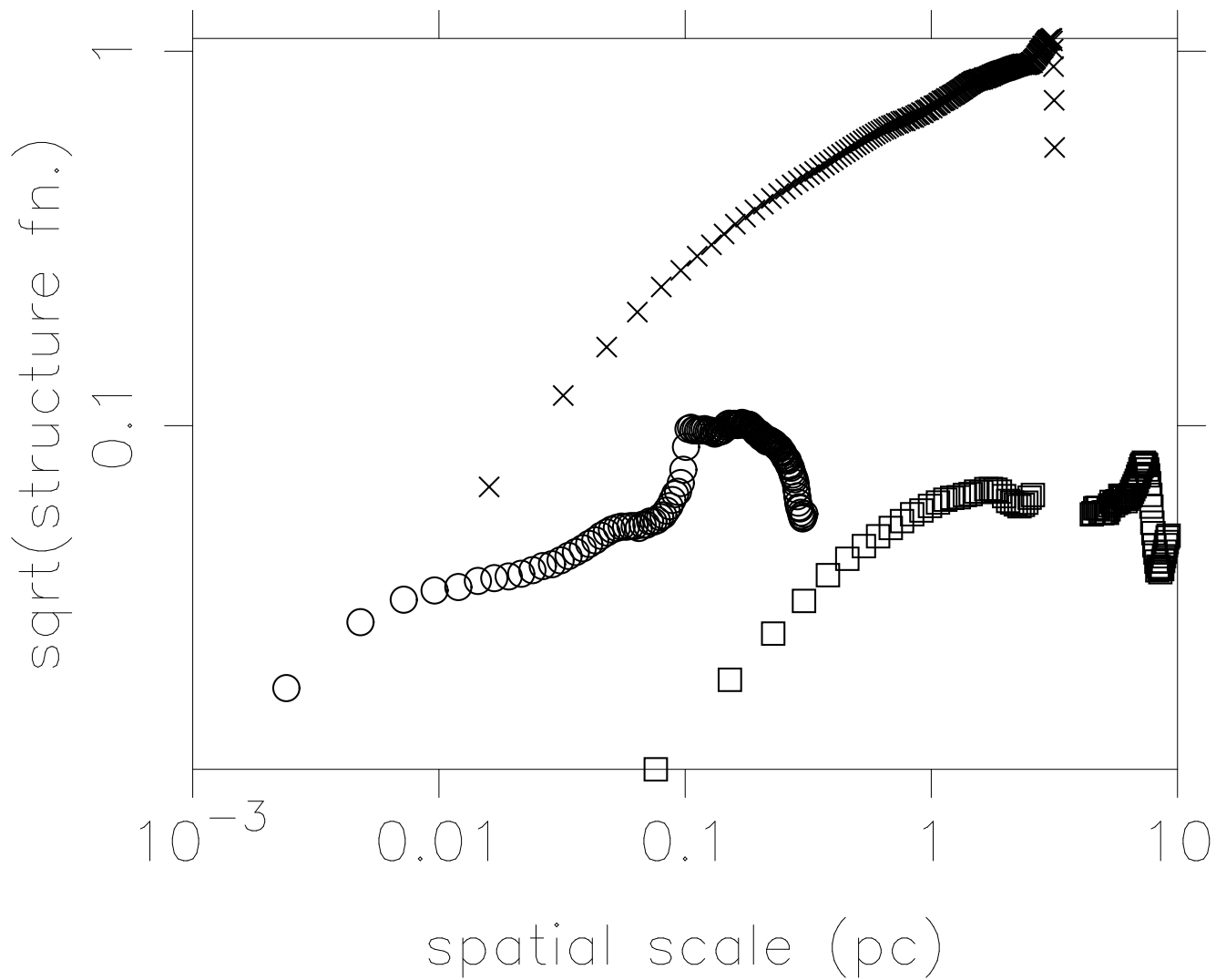}
\caption{The structure function towards Cas A and Cygnus A. The square root of the structure 
function (i.e., the r.m.s. value) of the optical depth distribution in the Perseus arm towards
Cas A (indicated by $\times$s) is compared with those towards Cygnus A in the Local (indicated by circles) 
and the Outer (indicated by squares) arms. The spatial scales correspond to the assumed distances of 
0.5, 2, and 15 kpc for the Local, the Perseus, and the Outer arms respectively. 
See Section 4 for further details.}
\end{figure*}

\subsection{Implications for the distribution of HI density and for pressure equilibrium}

The derived (CLEAN) power spectra shown in Fig. 6 represents p$_{\tau}(k)$, 
the spatial power spectrum 
of the optical depth distribution. The optical depth,
 $\tau$, is related to
the column density (N$_{H}$) and (density weighted harmonic) mean spin temperature (T$_{s}$) 
of the HI gas as $\tau \propto 
N_{H}/T_{s}$. The column density of HI is related to the volume density (n$_{H}$) as N$_{H}$ = 
n$_{H}$l, where, l is the line of sight path length of the HI gas. The pressure of
the HI gas (P) viewed locally is related to the volume density and kinetic temperature (T$_{k}$) as 
P$\propto$ n$_{H}$T$_{k}$, assuming it to be a perfect gas.  For the cold gas in the Galaxy, 
T$_{s} \approx$ T$_{k}$. If the different density regions are in pressure equilibrium
(i.e., n$_{H} \propto$ 1/T$_{k}$) with
each other then this implies that $\tau \propto$ n$_{H}^{2}$. In general, if the equation of
state for the cold HI gas is given by T$_{k} \propto$n$_{H}^{\gamma}$, then the optical 
depth will be proportional to n$_{H}^{1-\gamma}$. Pressure equilibrium corresponds to 
$\gamma$ = --1.  
It is not straightforward to predict the power law index of the density spectrum based on  
that of the observed optical depth.
However, it is possible to explore the relation between these two through suitable Monte Carlo 
simulations involving 
an assumed power law spectrum of density and a value of $\gamma$.  Our attempts to do so
have indicated that the slope of the computed power spectrum of opacity is nearly the same
as that of the assumed slope of the 3-d power spectrum for the HI density.  The two slopes
differ marginally if in the assumed density distribution, the density fluctuations far exceeded 
the mean density. Otherwise, the resultant slope showed no dependence on the assumed value
of $\gamma$, indicating that the assumption of pressure equilibrium is reasonable. 
 This independence of the slope on the assumed value of $\gamma$ is not surprising since any non-linear 
dependence of $\tau$ on $n_H$ implied  by the equation of state 
($\tau \propto$ n$_{H}^{1-\gamma}$) reduces to the first-order
to a dominant linear dependence ($\delta\tau \propto$ (1-$\gamma )\delta$n$_{H}$) if the
fluctuation in density ($\delta$n$_{H}$) is small compared to the mean density ($<$n$_{H}>$).
Hence, the nature of the observed spectrum of $\tau$ distribution must 
closely describe the corresponding spectrum of the HI density.

\section{DISCUSSION}

As can be seen in Fig. 6, the spectrum obtained in Case B (spectrum of the
 optical depth averaged over several velocity channels) 
has about a factor of 3 higher power than the Case A spectrum (average of the spectra of 
several velocity channels). This difference indicates that  
 there is partial correlation in structures as a function of velocity. The spectra are
normalized suitably such that if the structures
were fully correlated across N channels, we would observe an increase in power by a factor of
N. Hence, the observed increase by a factor of three in power indicates
a correlation scale of $\sim$ 2 km s$^{-1}$ in velocity. 
This correlation scale in velocity is expected since the typical value of
turbulence for the cold HI gas in the Galaxy is $\sim$ 4 km s$^{-1}$ (Radhakrishnan et al. 1972). 
%If this correlation is to be attributed entirely to longitudinally elongated
%structures, the implied elongation factors will be in excess of 100, a quite unlikely morphology. 

The power spectrum of density fluctuations in the cold atomic gas towards Cas A has a slope
of $\sim$ 2.75. Adopting a distance of 2 kpc for this absorbing gas, the linear
scales probed range between 0.07 and 3 pc, corresponding to the resolution (7$''$)
and the size of Cas A ($\sim$5$'$) respectively.
Compared to these transverse scales, the longitudinal (or, the line of sight) scales corresponding to each
of the channels (i.e., velocity resolution) must be significantly larger. Otherwise,  
the power spectrum obtained from the integrated image (Case B) 
should have a different (steeper) slope compared to that for a single channel image (Case A).
However, as has been demonstrated in Fig. 6, there is no significant difference in the two
slopes.  Thus, the opacity image 
in a single channel corresponds to a longitudinally integrated version of the underlying 
3-dimensional opacity distribution. Combining this result with the discussion in Section 4, the power 
spectrum represents the 3-d 
power spectrum of density fluctuations in the cold atomic gas over linear scales
of 0.07 to 3 pc. The HI emission analysis (Green 1993) provided a similar slope, 
over the larger scales of 50 to 200 pc. Assuming no changes in the slope of the power spectrum between 
these two ranges, a constant slope is implied over 3 orders of magnitude in linear scales. 
It is noteworthy that the estimated slope of 2.75 for the power spectrum differs significantly from
 the Kolmogorov slope (11/3 = 3.67) expected for the 3-dimensional power 
spectrum of a turbulence-dominated incompressible fluid.

In a recent paper, Lazarian, \& Pogosyan (2000) draw attention to the 
modification of the HI density spectrum due to the velocity field. They point out that
the velocity range needs to be {\it thick} in order to obtain the true density power spectrum. 
They estimate a power law index close to the Kolmogorov value for the power spectrum of density using
the  HI emission data of the Galaxy (Green 1993) and of the Small Megallanic Clouds (Stanimirovic 
et al. 1999).  
Being aware of this complication, we have obtained the power spectra before (Case A), and 
after (Case B) averaging in velocity (Section 4). Within the errors of estimation, the two slopes
agree (Fig. 6), indicating that the effect of averaging over velocity is not a substantial effect.

It may not be surprising that the slope of the power spectrum of density fluctuations
in the cold atomic gas in the Galaxy is different
from the Kolmogorov slope considering that : (a) the HI gas in the interstellar medium is compressible,
and, (b) the Kolmogorov index is expected if the energy input into the ISM at some large
scale cascades down to small scales through a turbulence mechanism. 
However, the exact nature and the scales at which the energy is input, and  its cascade
 to smaller scales are determined
by a range of processes in the interstellar medium such as the supernovae, and stellar winds.
Although a qualitative discussion of many such aspects is available (Lazarian 1995), 
the quantitative implications of these processes on the slope of the power spectrum is 
quite uncertain. 
%If we
%consider large-scale HI irregularities of size $\sim$ 1 pc, densities $\sim$ 20 cm$^{-3}$, 
%kinetic temperatures $\sim$ 100 K, and internal velocities
%$\sim$ a few km s$^{-1}$, it can be shown that the associated Reynolds number is high enough to 
%expect turbulence
%to have developed in such systems. However, what is not clear is the cascade of power in turbulence from
%larger to smaller scales. This is because supernova remnants play a dominant role in
%creating and destroying structures in HI gas. When an HI structure is overtaken by a shock 
%wave it would be compressed and consequently flattened. The various instabilities which will eventually
%set in will fragment the structure. The detailed evolution  depends on whether the
%shock propagation time scale across the structure is smaller or larger than the time scale for
%the evolution of the supernova remnant. A simple picture of large scale turbulence cascading to
%smaller scales is far from correct in this case. 

Another important implication of the derived power spectral description is to the long-standing
puzzle of the observed opacity variations on small transverse scales down to tens of AU. 
These opacity variations have been interpreted in the past as evidence for wide-spread small-scale
structure in the HI distribution with volume densities orders of magnitude higher than those
implied by the parsec-scale component. It has been recently emphasized (Deshpande 2000) that 
there is a need to recognize the nature of the actual quantity measured 
at small scales; almost all scales contribute to the measured opacity differences at small scales.
Furthermore, if the power spectrum with a slope of 2.75 estimated for the Cas A Perseus arm data
 is extrapolated
to these small scales, optical depth variations with an r.m.s. of 0.1 can be expected
on scales of $\sim$ 100 AU (Deshpande 2000). The recent HI absorption measurements using the 
Very Long Baseline Array and the Very Large Array indicate that 5 out of the 7 sources observed
show opacity variations consistent with this extrapolation (Faison et al. 1998, Faison \& Goss, 2000).
However, 3C138 \& 3C147 show optical depth fluctuations an order of magnitude larger over linear
sizes of 10 - 20 AU indicating real two dimensional structures in HI opacity but, note that the
associated spectra appear much shallower than in the Cas A case.

Our investigations in two different directions (Cas A, and Cygnus A) have shown both 
similarities and differences in the derived descriptions in terms of slopes, and powers associated
with the power spectra (and structure functions) of the HI distribution. 
For a relatively complete picture 
to emerge it is desirable that such studies are carried out in different directions in the Galaxy.

\acknowledgements

The analysis software used here was based largely on a package developed by Jayadev
Rajagopal with one of us (AAD). We would like to thank Rajaram Nityananda, and V. Radhakrishnan
 for useful discussions, and comments on the manuscript. National Radio Astronomy Observatory 
is a facility of the National Science Foundation operated under cooperative agreement by Associated
Universities, Inc.

\clearpage

\clearpage

\end{document}